\documentclass[%
 reprint,
 amsmath,amssymb,
 aps,
]{revtex4-2}
\usepackage{amsfonts}
\usepackage{amsmath}
\usepackage{amssymb}
\usepackage{charter}
\usepackage{subfigure}
\usepackage{dcolumn}
\usepackage{graphicx}
\usepackage{float}
\usepackage{graphicx}
\usepackage{dcolumn}
\usepackage{bm}

\begin{document}

\title{Noiseless linear amplification-based quantum Ziv-Zakai bound for
phase estimation and its Heisenberg error limits in noisy scenarios}
\author{Wei Ye$^{1}$}
\author{Peng Xiao$^{1}$}
\author{Xiaofan Xu$^{1}$}
\author{Xiang Zhu$^{1}$}
\author{Yunbin Yan$^{2}$}
\author{Lu Wang$^{2}$}
\author{Jie Ren$^{3}$}
\author{Yuxuan Zhu$^{1}$}
\author{Ying Xia$^{4}$}
\author{Xuan Rao$^{1}$}
\author{Shoukang Chang$^{1,5,6}$}
\thanks{Corresponding author. a2195662480@163.com}

\affiliation{$^{{\small 1}}$\textit{School of Information Engineering, Nanchang Hangkong
University, Nanchang 330063, China}\\
$^{{\small 2}}$\textit{Shijiazhuang Campus, Army Engineering University of
PLA,Shijiazhuang 050035, China}\\
$^{{\small 3}}$\textit{Taiyuan Satellite Launch Center,  Taiyuan
030045, China}\\
$^{{\small 4}}$\textit{School of Physics, Sun Yat-sen University, Guangzhou
510275, China}\\
$^{{\small 5}}$\textit{MOE Key Laboratory for Nonequilibrium Synthesis and
Modulation of Condensed Matter, Shaanxi Province Key Laboratory of Quantum
Information and Quantum Optoelectronic Devices, School of Physics, Xi'an
Jiaotong University, Xi'an 710049, People's Republic of China}\\
$^{{\small 6}}$\textit{Dipartimento di Fisica \textquotedblleft Aldo
Pontremoli\textquotedblright , Universit\`{a} degli Studi di Milano, 20133
Milan, Italy}}

\begin{abstract}
In this work, we address the central problem about how to effectively find
the available precision limit of unknown parameters. In the framework of the
quantum Ziv-Zakai bound (QZZB), we employ noiseless linear amplification
(NLA)techniques to an initial coherent state (CS) as the probe state, and
focus on whether the phase estimation performance is improved significantly
in noisy scenarios, involving the photon-loss and phase-diffusion cases.
More importantly, we also obtain two kinds of Heisenberg error limits of the
QZZB with the NLA-based CS in these noisy scenarios, making comparisons with
both the Margolus-Levitin (ML) type bound and the Mandelstam-Tamm (MT) type
bound. Our analytical results show that in cases of photon loss and phase
diffusion, the phase estimation performance of the QZZB can be improved
remarkably by increasing the NLA gain factor. Particularly, the improvement
is more pronounced with severe photon losses. Furthermore in minimal photon
losses, our Heisenberg error limit shows better compactness than the cases
of the ML-type and MT-type bounds. Our findings will provide an useful
guidance for accomplishing more complex quantum information processing tasks.%
\end{abstract}

\maketitle
\section{Introduction}

As an important pillar of science and technology, quantum metrology aims to
precisely measure physical quantities within the framework of quantum
theory, with applications ranging from gravitational wave detection \cite{1}
to atomic-scale structural analysis \cite{2}. For this purpose, there are
two commonly used strategies \cite{3,4,5,6,7,8,9}, i.e., improving the
precision of parameter estimation in optical interferometers and pursuing
high precision limit with the help of the quantum Fisher information (QFI).
For the former, non-Gaussian operations \cite{10,11,12} and noiseless linear
amplifications (NLA) \cite{13,14} are often used, which play a vital role in
quantum communication \cite{15,16} and quantum steering \cite{17,18}. In
particular, G.S. Agarwal and L. Davidovich applied the quantum-amplification
strategy into the SU(1,1) and atomic interferometers, and then presented how
their scheme results in the Heisenberg-scaling precision of phase estimation
\cite{19}. For the latter, it is common to exploit the famed quantum Cram%
\'{e}r-Rao bound (QCRB) to quantify the minimum value for the mean square
error in quantum metrology \cite{20,21,22,23,24}. Generally, the QCRB can be
described by the QFI, which is helpful to solve the problem of single- and
multi-parameter estimations, so as to enhance its practical application
value \cite{25}. For instance, with the help of the QFI matrix in the
multi-parameter estimation, the networked quantum sensors can improve the
estimation precision \cite{26,27}. For all this, when the likelihood
function is extremely non-Gaussian or under the limit of trials, the QCRB
appears to be in a serious underestimation of estimation precision \cite%
{28,29,30}.
\begin{figure*}[tbp]
\label{Fig1} \centering \includegraphics[width=1.3\columnwidth]{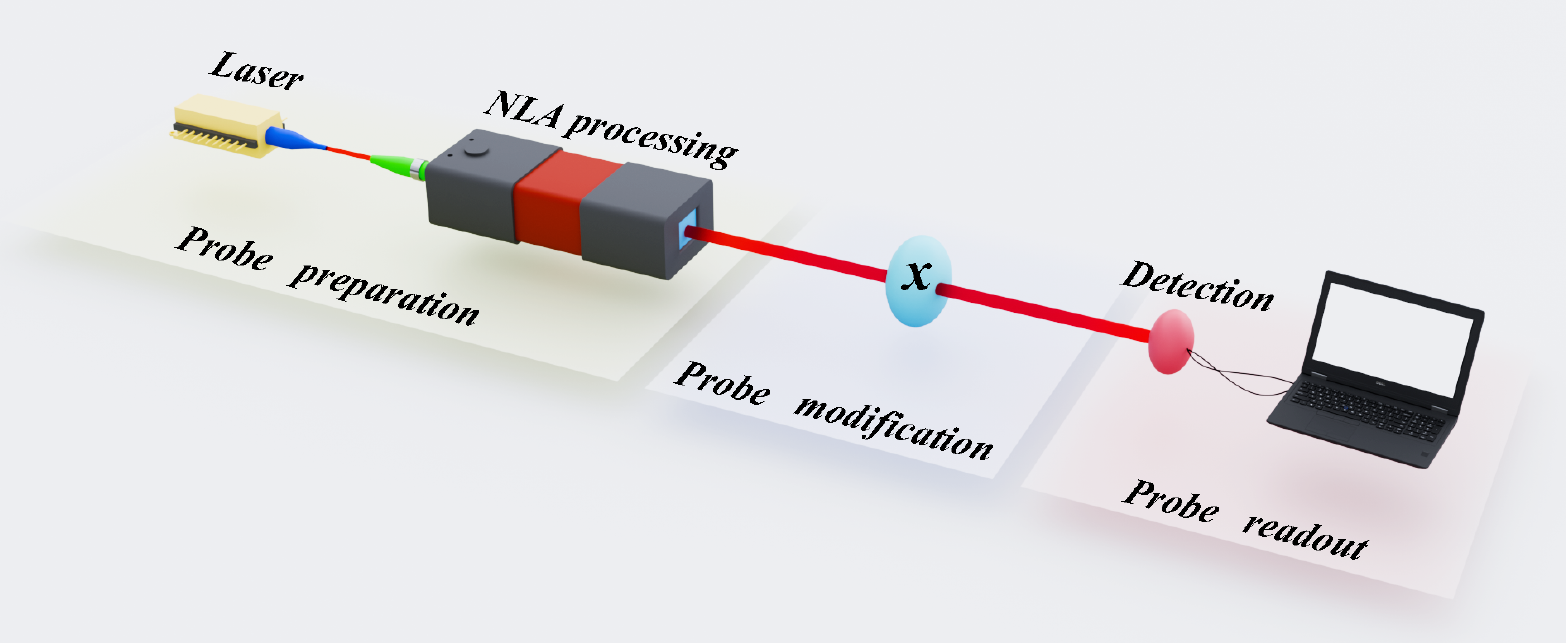}
\caption{{}(Color online) The setup of the QZZB with the NLA-based CS, where
the initial CS generated by the laser is amplified by using the NLA
processing, then interfered with the phase shifter $x$, and finally detected
by using the POVM in order to evaluating phase estimation precision of the
QZZB. }
\end{figure*}

To solve the aforementioned problem, the quantum Ziv-Zakai bound (QZZB) has
emerged as a promising candidate, which connects the mean square error in
quantum parameter estimation to the error probability in quantum hypothesis
testing \cite{28,31,32,33,34,35,50,51,52}. By applying the QZZB into optical
phase estimations, M. Tsang directly showed the tighter advantages of the
QZZB over the QCRB \cite{28}. Based on the aforementioned work \cite{28}, Y.
R. Zhang and H. Fan extended the QZZB to multiple-parameter case and then
derived two kinds of quantum metrological bounds in the noisy system \cite%
{36}. In addition, S. K. Chang \textit{et al.} further presented the QZZB
for phase estimation in the frameworks of both photon loss and phase
diffusion \cite{37}, finding that the QZZB for an initial coherent state
(CS) can show the best estimation performance in the both noisy scenarios.
In this context, it is natural to ask whether the NLA process improves the
phase estimation performance of the QZZB in the noisy environment, which has
never been discussed and studied before.

In this paper, by employing the NLA to an initial CS, we examine the phase
estimation performance of the QZZB with the photon-loss and phase-diffusion
scenarios, and then derive two Heisenberg error limits of the QZZB with the
NLA-based CS in noisy scenarios, also making comparisons with both the
Margolus-Levitin (ML) type bound and the Mandelstam-Tamm (MT) type bound
\cite{36}. Our results indicate that for photon-loss and phase-diffusion
cases, the corresponding phase estimation performance of the QZZB can be
enhanced effectively by increasing the NLA gain factor. Particularly, the
more serious the photon losses, the more obvious the improvement effect.
Especially for small photon loss cases, Heisenberg error limits of our
scheme show the best compactness when compared to ML-type and MT-type bounds.

The structure of this paper is as follows. For photon-loss and
phase-diffusion scenarios, respectively corresponding to Sections II and
III, we present the derivation of the QZZB with the NLA-based CS by using
the variational method. In Section IV, we also pay attention to deriving two
Heisenberg error limits of the QZZB with the NLA-based CS in photon-loss and
phase-diffusion scenarios, respectively. In Section V, we give the
performance analyses of the QZZB with the NLA-based CS in noisy scenarios.
Finally, we summarize our results.

\section{The QZZB with the NLA-based CS in photon-loss scenarios}

In this section, we turn our attention on the derivation of the QZZB with
the NLA-based CS in photon-loss scenarios by means of the variational
method. For this purpose, we shall divide it into the following three steps.

The first step is that we need to recall the existing results of the QZZB in
an ideal scenario. In quantum parameter-estimated theory, the observation
density can be expressed as%
\begin{equation}
p\left( y|x\right) =\text{Tr}[\hat{\rho}\left( x\right) \Pi (y)],  \label{1}
\end{equation}%
where $\hat{\rho}\left( x\right) $ is the density operator as a function of
the unknown phase parameter $x$,\ $\Pi (y)$ is the positive operator-valued
measure (POVM) viewed as the measurement model, and the symbol of Tr is the
operator trace. In this situation, as given in Refs. \cite{28,36}, the lower
bound of the minimum error probability (denoted as $\Pr\nolimits_{e}^{el}%
\{x,x+\tau \}$) is written as
\begin{eqnarray}
\Pr\nolimits_{e}^{el}(x,x+\tau ) &\geq &\frac{1}{2}(1-\frac{1}{2}\left\Vert
\hat{\rho}\left( x\right) -\hat{\rho}\left( x+\tau \right) \right\Vert _{1})
\notag \\
&\geq &\frac{1}{2}[1-\sqrt{1-F\{\hat{\rho}\left( x\right) ,\hat{\rho}\left(
x+\tau \right) \}}],  \label{2}
\end{eqnarray}%
with $\left\Vert \hat{O}\right\Vert _{1}$=Tr$\sqrt{\hat{O}^{\dagger }\hat{O}}
$ and the so-called Uhlmann fidelity F\{$\rho \left( x\right) ,\rho \left( x%
\text{+}\tau \right) $\}=[Tr$\sqrt{\sqrt{\rho \left( x\right) }\rho \left( x%
\text{+}\tau \right) \sqrt{\rho \left( x\right) }}$]$^{2}$. When assuming
that the quantum state $\hat{\rho}\left( x\right) $ carries an unknown phase
parameter $x$, which satisfies the following unitary evolution%
\begin{equation}
\hat{\rho}\left( x\right) =e^{-i\hat{H}x}\hat{\rho}_{in}e^{i\hat{H}x},
\label{3}
\end{equation}%
with the initial state $\hat{\rho}_{in}$ and the effective Hamiltonian
operator $\hat{H}$, one can thus obtain the inequality $F(\hat{\rho}\left(
x\right) ,\hat{\rho}\left( x+\tau \right) )\geq \left\vert \text{Tr}(\hat{%
\rho}_{in}e^{-i\hat{H}\tau })\right\vert ^{2}$ \cite{41}. Similar to derive
the classical Ziv-Zakai bound method \cite{28}, we also need to take an
assumption that the prior probability density $p(x)$ is the uniform window
with the mean $\zeta $ and the width $y$ given by%
\begin{equation}
p(x)=\frac{1}{y}rect\left( \frac{x-\zeta }{y}\right) ,  \label{4}
\end{equation}%
and then leave out the so-called optional valley-filling operation, which
renders the bound tighter. Thus, the QZZB for an ideal scenario is finally
described as \cite{17}
\begin{equation}
\sum \geq \sum \nolimits_{Z}=\int_{0}^{y}d\tau \frac{\tau }{2}\left( 1-\frac{%
\tau }{y}\right) [1-\sqrt{1-F(\tau )}],  \label{5}
\end{equation}%
where $F(\tau )\equiv \left \vert \text{Tr}(\hat{\rho}_{in}e^{-i\hat{H}\tau
})\right \vert ^{2}$ is the lower bound of the fidelity $F\{ \rho \left(
x\right) ,\rho \left( x+\tau \right) \}$.

The second step is that for practical relevance the derivation of the QZZB
is in photon-loss scenarios because the encoding process of the initial
state $\hat{\rho}_{in}$ to the unknown phase parameter $x$ is inevitably
influenced by the surrounding environment. It is worthy noticing that in
photon-loss scenarios, such an encoding process is not a unitary evolution,
which results in the fact that Eq. (\ref{5}) cannot be used directly to
derive the QZZB. Fortunately, Escher \textit{et al}. proposed an variational
method for achieving the upper bound in noisy environment \cite{38,39},
where the encoding process converts to the unitary evolution $\hat{U}%
_{S+E}\left( x\right) $ by adding the degree of freedom acting as an
environment $E$ for the probe system $S$.

Based on the aforementioned method, for an initial pure state $\hat{\rho}%
_{S}=\left\vert \psi _{S}\right\rangle \left\langle \psi _{S}\right\vert $
in the probe system $S,$ we should expand the size of Hilbert space $S$
together with the photon-loss environment space $E$, so that the initial
pure state after going through the unitary evolution $\hat{U}_{S+E}\left(
x\right) $ can be described as
\begin{eqnarray}
\hat{\rho}_{S+E}\left( x\right) &=&\hat{U}_{S+E}\left( x\right) \hat{\rho}%
_{S}\otimes \hat{\rho}_{E_{0}}\hat{U}_{S+E}^{\dag }\left( x\right)  \notag \\
&=&\sum_{k=0}^{\infty }\hat{\Pi}_{k}\left( x\right) \hat{\rho}_{S}\otimes
\hat{\rho}_{E_{k}}\hat{\Pi}_{k}^{\dagger }\left( x\right) ,  \label{6}
\end{eqnarray}%
where $\hat{\rho}_{E_{0}}=\left\vert 0_{E}\right\rangle \left\langle
0_{E}\right\vert $ represents the initial state in photon-loss environment
space $E$, $\hat{\rho}_{E_{k}}=\left\vert k_{E}\right\rangle \left\langle
k_{E}\right\vert $ represents the orthogonal basis of $\hat{\rho}_{E_{0}},$
and $\hat{\Pi}_{k}\left( x\right) $ represents the Kraus operator given by
\begin{equation}
\hat{\Pi}_{k}\left( x\right) =\sqrt{\frac{\left( 1-\eta \right) ^{k}}{k!}}%
e^{-ix\left( \hat{n}-\mu k\right) }\eta ^{\frac{\hat{n}}{2}}\hat{a}^{k},
\label{7}
\end{equation}%
with a photon-loss strength $\eta $, a variational parameter $\mu ,$ and a
photon number operator $\hat{n}=\hat{a}^{\dag }\hat{a}$. Among them, $\eta
=0 $ and $1$ are the complete-absorption and ideal cases, respectively. By
using the Uhlmann's theorem \cite{40}, the fidelity for photon-loss
scenarios in the enlarged system $S+E$ can be written as
\begin{equation}
F_{L_{1}}(\tau )=\max_{\left\{ \hat{\Pi}_{k}\left( x\right) \right\}
}F_{Q_{1}}\{\hat{\rho}_{S+E}\left( x\right) ,\hat{\rho}_{S+E}\left( x+\tau
\right) \},  \label{9}
\end{equation}%
with
\begin{eqnarray}
&&F_{Q_{1}}\{\hat{\rho}_{S+E}\left( x\right) ,\hat{\rho}_{S+E}\left( x+\tau
\right) \}  \notag \\
&=&\left\vert \left\langle \psi _{S+E}\left( x\right) |\psi _{S+E}\left(
x+\tau \right) \right\rangle \right\vert ^{2}  \notag \\
&=&\left\vert \text{Tr}(\hat{\rho}_{S}\left[ \eta e^{-i\tau }+\left( 1-\eta
\right) e^{i\tau \mu }\right] ^{\hat{n}})\right\vert ^{2},  \label{10}
\end{eqnarray}%
corresponding to the lower bound of the fidelity in photon-loss scenarios.
For the sake of expression, we take $F_{Q_{1}}\left( \tau \right) \equiv
F_{Q_{1}}\{\hat{\rho}_{S+E}\left( x\right) ,\hat{\rho}_{S+E}\left( x+\tau
\right) \}$ as follows. The more details for the derivation can be seen in
Ref. \cite{37}. Thus, by taking the optimal variational parameter $\mu
_{opt} $, the maximum value of the fidelity in photon-loss scenarios given
in\ Eq. (\ref{10}) can be achieved. In this case, Eq. (\ref{2}) under the
photon losses (denoted as $\Pr\nolimits_{e_{L_{1}}}^{el}(x,x+\tau )$) can be
rewritten as
\begin{eqnarray}
\Pr\nolimits_{e_{L_{1}}}^{el}(x,x+\tau ) &\geq &\frac{1}{2}(1-\frac{1}{2}%
\left\Vert \hat{\rho}_{S+E}\left( x\right) +\hat{\rho}_{S+E}\left( x+\tau
\right) \right\Vert _{1})  \notag \\
&\geq &\frac{1}{2}[1-\sqrt{1-F_{L_{1}}(\tau )}].  \label{11}
\end{eqnarray}%
Similar to obtain Eq. (\ref{5}), according to Eq. (\ref{11}), the QZZB in
photon-loss scenarios is given by
\begin{equation}
\sum\nolimits_{L_{1}}\geq \sum\nolimits_{Z_{L_{1}}}^{PL}=\int_{0}^{y}\tilde{F%
}_{1}(\tau )d\tau ,  \label{12}
\end{equation}%
with the superscript PL being the phase-loss case, and the photon-loss
generalized fidelity $\tilde{F}_{1}(\tau )=\sin (\pi \tau /y)F_{L_{1}}(\tau
)y/16$, where we have used the following inequalities \cite{28}
\begin{eqnarray}
1-\sqrt{1-F_{L_{1}}(\tau )} &\geq &\frac{F_{L_{1}}(\tau )}{2},  \notag \\
\tau \left( 1-\frac{\tau }{y}\right) &\geq &\frac{y}{4}\sin \frac{\pi \tau }{%
y},  \label{13}
\end{eqnarray}%
with $F_{L_{1}}(\tau )$ satisfying the conditions of $0\leq F_{L_{1}}(\tau
)\leq 1$ and $0\leq \tau \leq y$.

The last step is that we derive the QZZB with the NLA-based CS in photon
losses, as shown in Fig. 1. Theoretically, an ideal noiseless linear
amplifier can be implemented by the operator $g^{\hat{a}^{\dag }\hat{a}}$
(the amplification factor: $g>1$), which satisfies the following
transformation
\begin{equation}
\left\vert \gamma \right\rangle =\frac{1}{\sqrt{P_{d}}}g^{\hat{a}^{\dag }%
\hat{a}}\left\vert \alpha \right\rangle =\frac{1}{\sqrt{P_{d}}}\exp \{\left(
g^{2}-1\right) |\gamma |^{2}/2\}\left\vert \gamma \right\rangle ,  \label{14}
\end{equation}%
where $P_{d}=\exp \{\left( g^{2}-1\right) |\gamma |^{2}\}$ is the
normalization factor, $\left\vert \alpha \right\rangle $ is the initial
input CS and $\left\vert \gamma \right\rangle $ is the probe state with $%
\gamma =g\alpha $, which is called as the NLA-based CS. By using Eq. (\ref%
{10}), the lower bound of the fidelity with the NLA-based CS $\left\vert
\gamma \right\rangle $ in photon-loss scenarios is thus given by%
\begin{equation}
F_{Q_{1}}\left( \tau \right) _{\left\vert \gamma \right\rangle }=\exp \left[
2N_{\gamma }\left( \eta \cos \tau +\left( 1-\eta \right) \cos \tau \mu
-1\right) \right] ,  \label{15}
\end{equation}%
where $N_{\gamma }=g^{2}N_{\alpha }$ related to the average photon number of
the initial CS $\left\vert \alpha \right\rangle $. By taking $\mu _{opt}=0$
that maximizes the lower bound, Eq. (\ref{15}) can be further written as
\begin{equation}
F_{L_{1}}(\tau )_{\left\vert \gamma \right\rangle }=\exp \left[ 2\eta
N_{\gamma }\left( \cos \tau -1\right) \right] .  \label{16}
\end{equation}%
Finally, by seting $y=2\pi $,\ combining Eqs. (\ref{12}) and (\ref{16}),\
the QZZB with the NLA-based CS $\left\vert \gamma \right\rangle $ in
photon-loss scenarios is calculated as
\begin{equation}
\sum\nolimits_{Z_{L_{1}}}\geq \sum\nolimits_{Z_{L_{1}}\gamma }^{PL}=\frac{%
\pi ^{3/2}e^{-4\eta N_{\gamma }}}{8\sqrt{\eta N_{\gamma }}}\text{erfi}(2%
\sqrt{\eta N_{\gamma }}).  \label{17}
\end{equation}%
From Eq. (\ref{17}), when $g=1,$ the QZZB with the NLA-based CS can reduce
the one with the initial CS denoted as $\sum\nolimits_{Z_{L_{1}}\alpha }^{PL}
$.
\begin{figure}[tbp]
\label{Fig2} \centering \includegraphics[width=\columnwidth]{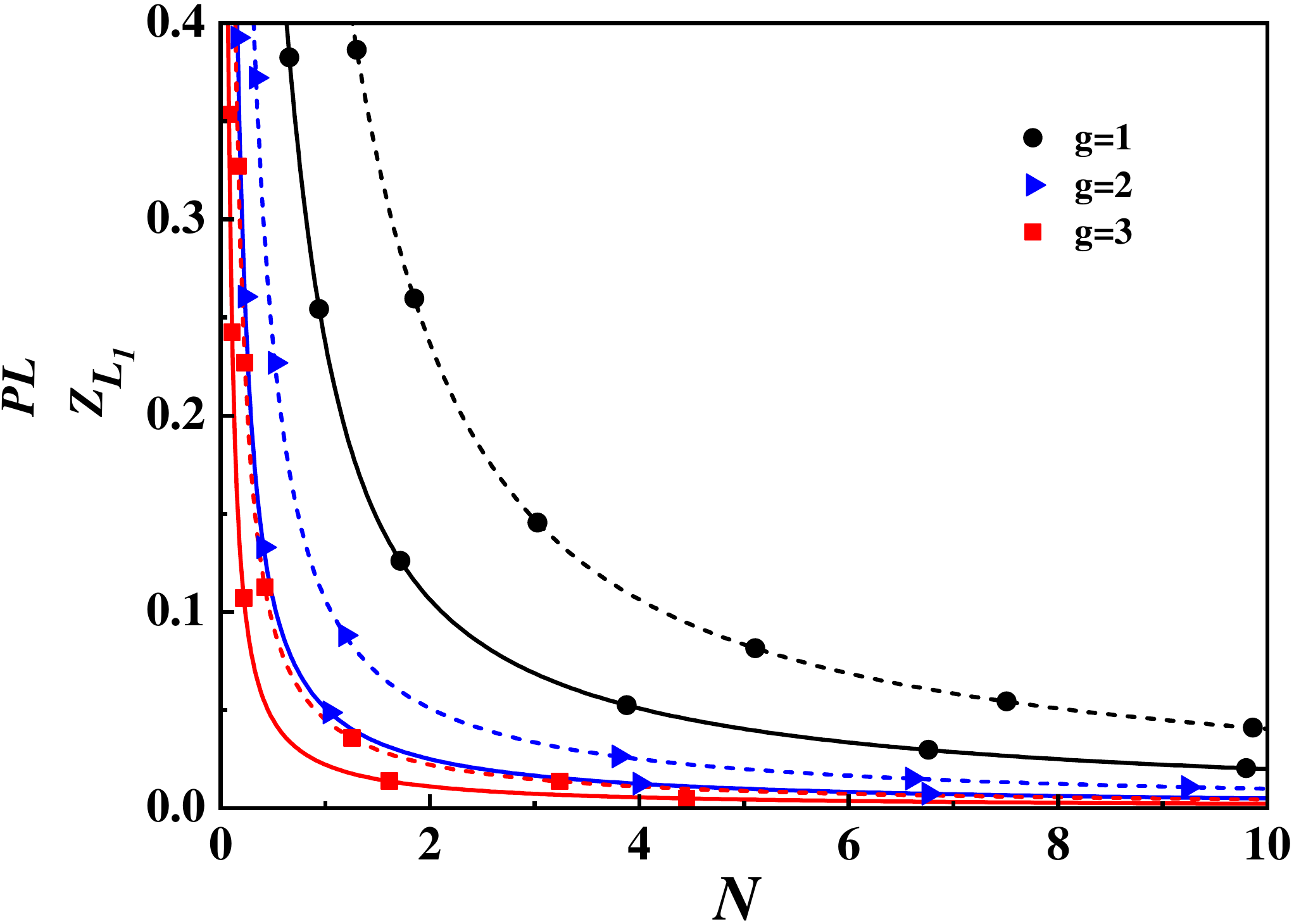}
\caption{{}(Color online) The QZZB $\sum\nolimits_{Z_{L_{1}}\protect\gamma %
}^{PL}$ with the NLA-based CS as a function of the total average photon
number $N$ for several different amplification factors $g=1,2,3$ when given
the photon-loss strength $\protect\eta =0.5$ (dashed lines) and the ideal
case $\protect\eta =1$ (solid lines). \ }
\end{figure}
\begin{figure}[tbp]
\label{Fig3} \centering \includegraphics[width=\columnwidth]{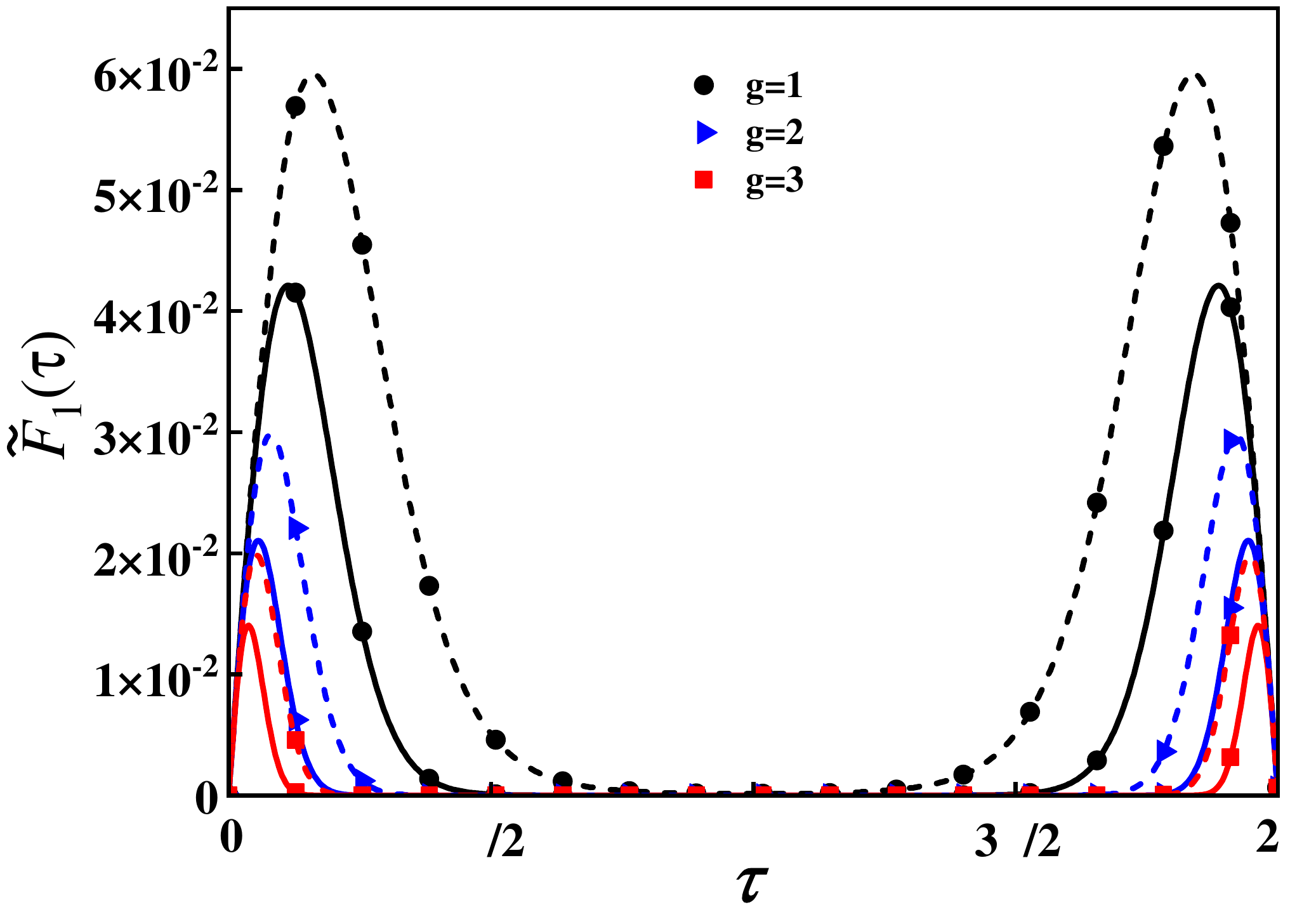}
\caption{{}(Color online) The photon-loss generalized fidelity $\tilde{F}%
_{1}(\protect\tau )$ changing with the phase difference $\protect\tau $ for
several different amplification factors $g=1,2,3$ when given the photon-loss
strength $\protect\eta =0.5$ (dashed lines), the ideal case $\protect\eta =1$
(solid lines) and $N=4.$\ }
\end{figure}
\begin{figure*}[tbp]
\label{Fig4} \centering
\subfigure[]{
\centering
\includegraphics[width=0.7\columnwidth]{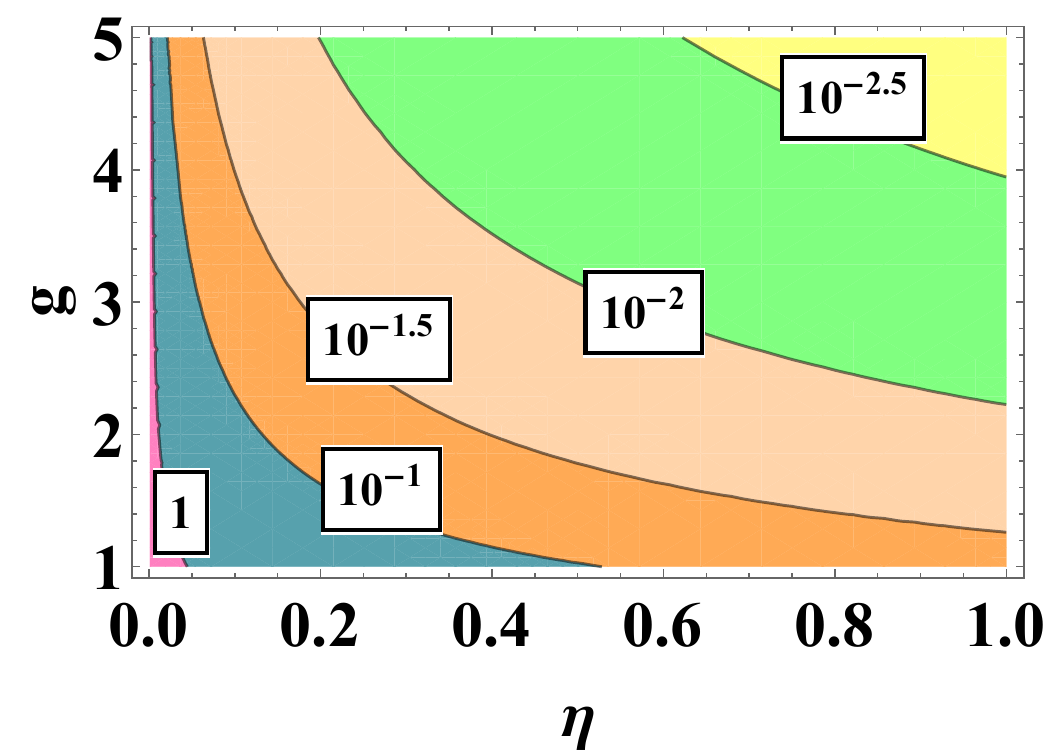}
\label{Fig4a}
}
\subfigure[]{
\includegraphics[width=0.7\columnwidth]{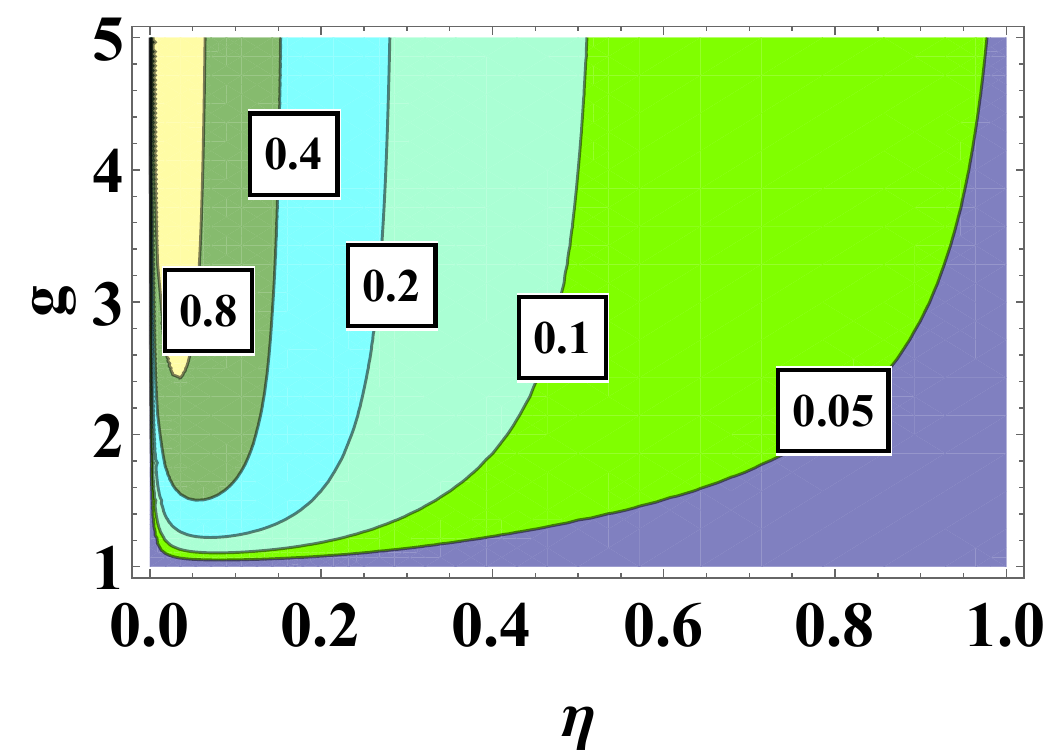}
\label{Fig4b}
}
\caption{{}(Color online) The QZZB (a) and (b) the differences of the QZZB
between without and with the ideal NLA, i.e., $\sum\nolimits_{Z_{L_{1}}%
\protect\alpha }^{PL}-$ $\sum\nolimits_{Z_{L_{1}}\protect\gamma }^{PL}$,
changing with the gain factor $g$ and the photon-loss strength $\protect\eta
$ when given $N=4$.}
\end{figure*}
\begin{figure*}[tbp]
\label{Fig5} \centering
\subfigure[]{
\centering
\includegraphics[width=0.95\columnwidth]{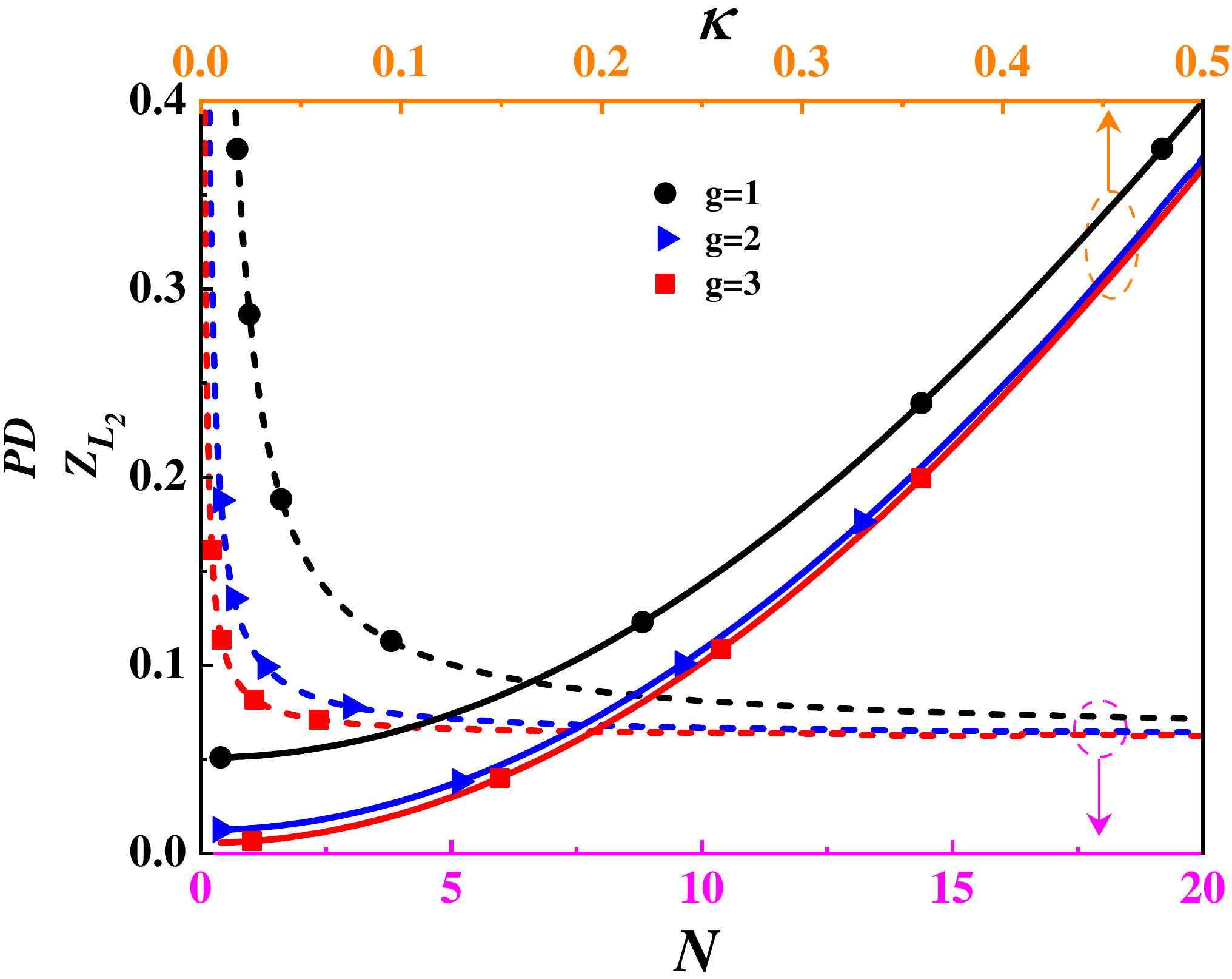}
\label{Fig5a}
}
\subfigure[]{
\includegraphics[width=\columnwidth]{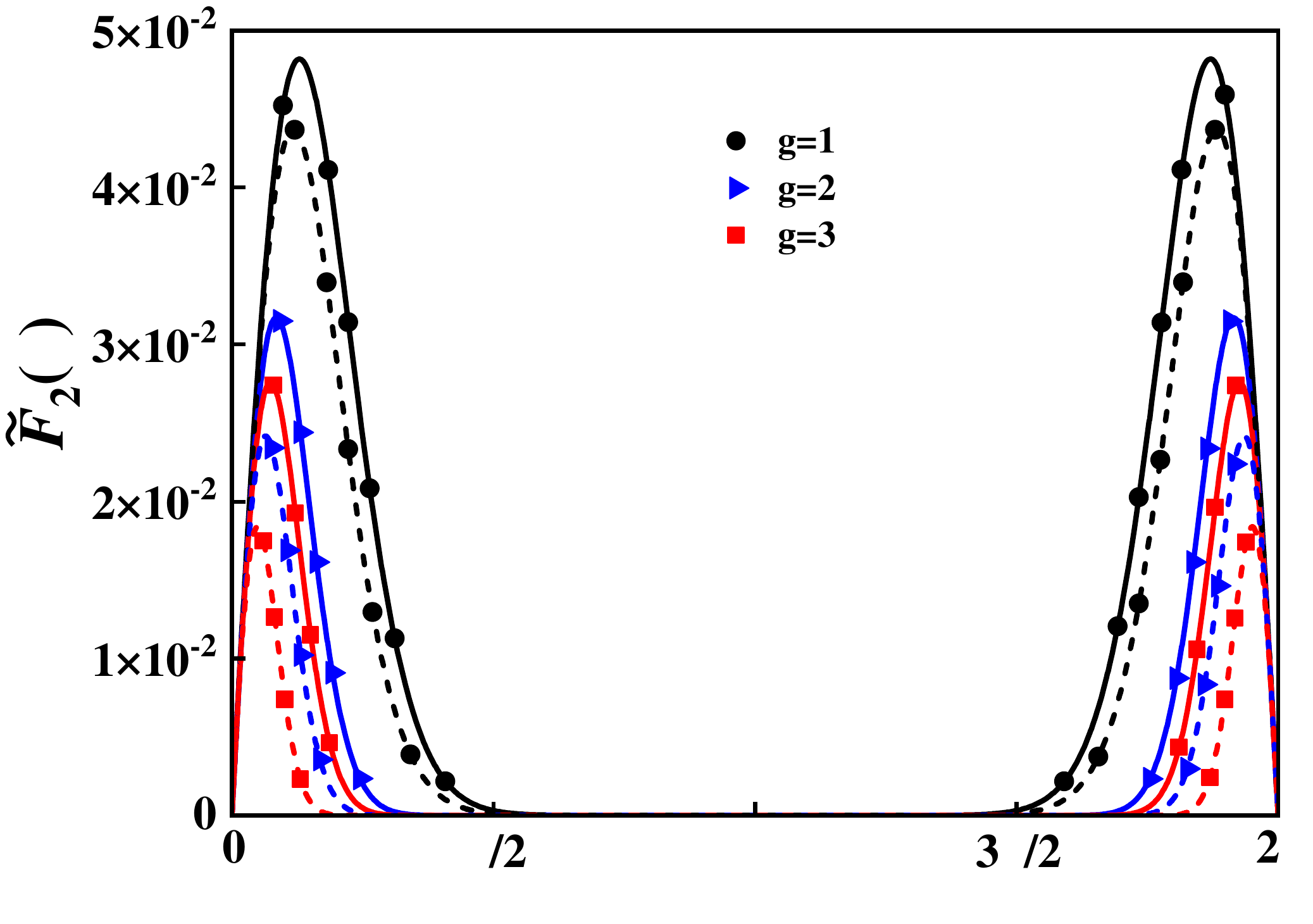}
\label{Fig5b}
}
\caption{{}(Color online) (a) The QZZB in photon-diffusion scenarios as a
function of the average photon number $N$ (lower coordinate) at a fixed $%
\protect\kappa =0.2$ and the photon-diffusion strength $\protect\kappa $
(upper coordinate) at a fixed $N=4$ for several different gain factors $%
g=1,2,3$. (b) The photon-diffusion generalized fidelity $\tilde{F}_{2}(%
\protect\tau )$ at a fixed $N=4.$\ changing with the phase difference $%
\protect\tau $ for several different amplification factors $g=1,2,3$ when
given photon-diffusion strengths $\protect\kappa =0.05$(dashed lines) and $%
0.1$ (solid lines). }
\end{figure*}

\section{The QZZB with the NLA-based CS in phase-diffusion scenarios}

So far, we have constructed the fundamental framework of the QZZB with the
NLA-based CS in photon-loss scenarios. In this section, we shall take the
impacts of phase diffusion on the phase estimation performance of the QZZB
into account. The reason is that phase diffusion is also the common noisy
environment, which has been investigated in the evaluation of the QCRB for
phase estimation \cite{39}.

As discussed in Refs. \cite{37,38,39}, the phase diffusion is used for
describing the interaction between the probe system $S$ and the environment $%
E$, which is given by the operator $\exp (i2\kappa \hat{n}\hat{q}_{E})$ with
the phase-diffusion strength of $\kappa $ and the dimensionless position
operator of the mirror $\hat{q}_{E}=(\hat{a}_{E^{\prime }}+\hat{a}%
_{E^{\prime }}^{\dagger })/\sqrt{2}$. Similar to Eq. (\ref{6}), the density
operator $\hat{\rho}_{S+E^{\prime }}^{PD}\left( x\right) $ in the enlarged
system $S+E$ can be described as%
\begin{eqnarray}
\hat{\rho}_{S+E}^{PD}\left( x\right) &=&\left\vert \Phi
_{S+E}(x)\right\rangle \left\langle \Phi _{S+E}(x)\right\vert  \notag \\
&=&\hat{U}_{S+E^{\prime }}^{PD}\left( x\right) \hat{\rho}_{S}\otimes \hat{%
\rho}_{E_{0}}^{PD}\hat{U}_{S+E^{\prime }}^{PD\dagger }\left( x\right) ,
\label{18}
\end{eqnarray}%
with the superscript PD being the phase-diffusion case throughout the whole
paper, $\hat{U}_{S+E^{\prime }}^{PD}\left( x\right) =\exp \{-i\hat{n}%
(x-2\kappa \hat{q}_{E})\}$ being the unitary operator of the enlarged system
$S+E$, and $\hat{\rho}_{E_{0}}^{PD}=\left\vert 0_{E}\right\rangle
\left\langle 0_{E}\right\vert $. Based on Eq. (\ref{18}), by introducing a
variational parameter $\mu $, the purified unitary evolution under the
asymptotic condition of $\sqrt{2}\kappa n\gg 1$ is given by
\begin{equation}
\widehat{\overline{\rho }}_{S+E}^{PD}\left( x\right) =\hat{u}_{E}^{PD}(x)%
\hat{\rho}_{S+E}^{PD}\left( x\right) \hat{u}_{E}^{PD\dagger }(x),  \label{19}
\end{equation}%
with $\hat{u}_{E}^{PD}(x)=\exp (ix\mu \hat{p}_{E}^{PD}/2\kappa )$ and $\hat{p%
}_{E}^{PD}=(\hat{a}_{E}-\hat{a}_{E}^{\dagger })/i\sqrt{2}$ representing the
dimensionless momentum operator of the mirror. Likewise, according to the
Uhlmann's theorem \cite{41}, in this situation, the fidelity in
phase-diffusion scenarios is expressed as
\begin{equation}
F_{L_{2}}(\tau )=\max_{\left\{ \left\vert \Phi _{S+E}(x)\right\rangle
\right\} }F_{Q_{2}}\left( \tau \right) ,  \label{20}
\end{equation}%
with $F_{Q_{2}}\left( \tau \right) \equiv F_{Q_{2}}\{\widehat{\overline{\rho
}}_{S+E}^{PD}\left( x\right) ,\widehat{\overline{\rho }}_{S+E}^{PD}\left(
x\right) \}$ representing the lower bound of the fidelity in phase-diffusion
scenarios, which can be further calculated as
\begin{equation}
F_{Q_{2}}\left( \tau \right) =\varphi (\kappa ,\tau ,\mu )\left\vert \text{%
Tr[}\hat{\rho}_{S}\exp \{i\tau \left( \mu -1\right) \hat{n}\}\text{]}%
\right\vert ^{2},  \label{21}
\end{equation}%
where $\varphi (\kappa ,\tau ,\mu )=e^{-\tau ^{2}\mu ^{2}/8\kappa ^{2}}$.
Combining Eqs. (\ref{20}) and (\ref{21}), the fidelity $F_{L_{2}}(\tau )$ in
phase-diffusion scenarios can be obtained by taking the optimal variational
parameter $\mu $. Likewise, in this situation, we can further obtain the
lower limit of the minimum error probability $\Pr%
\nolimits_{e_{L_{2}}}^{PD}(x,x+\tau )$ in phase-diffusion scenarios, i.e.,
\begin{eqnarray}
\Pr\nolimits_{e_{L_{2}}}^{PD}(x,x+\tau ) &\geq &\frac{1}{2}(1-\frac{1}{2}%
\left\Vert \widehat{\overline{\rho }}_{S+E}^{PD}\left( x\right) +\widehat{%
\overline{\rho }}_{S+E}^{PD}\left( x+\tau \right) \right\Vert _{1})  \notag
\\
&\geq &\frac{1}{2}[1-\sqrt{1-F_{L_{2}}(\tau )}].  \label{22}
\end{eqnarray}%
Thus, according to Eq. (\ref{22}) and setting $y=2\pi $, the QZZB for
phase-diffusion scenarios can be derived as%
\begin{equation}
\sum\nolimits_{L_{2}}\geq \sum\nolimits_{Z_{L_{2}}}^{PD}=\int_{0}^{y}\tilde{F%
}_{2}(\tau )d\tau ,  \label{23}
\end{equation}%
where $\tilde{F}_{2}(\tau )=y\sin (\pi \tau /y)F_{L_{2}}(\tau )/16$ is also
viewed as the phase-diffusion generalized fidelity. For the NLA-based CS,
the corresponding generalized fidelity can be calculated as
\begin{equation}
\tilde{F}_{2}(\tau )=\frac{\pi }{8}F_{L_{2}}(\tau )\sin (\tau /2),
\label{24}
\end{equation}%
with $F_{L_{2}}(\tau )$=$\max_{\mu }$\{$\varphi (\kappa ,\tau ,\mu
)e^{-2N_{\gamma }\text{(}1\text{-}\cos \text{[}\tau \text{(}\mu \text{-}1%
\text{)])}}$\}, so that
\begin{equation}
\sum\nolimits_{L_{2}}\geq \sum\nolimits_{Z_{L_{2}}\gamma }^{PD}=\int_{0}^{y}%
\tilde{F}_{2}(\tau )d\tau .  \label{25}
\end{equation}

\section{Heisenberg error limit of the QZZB with the NLA-based CS in noisy
scenarios}

\begin{figure}[tbp]
\label{Fig6} \centering \includegraphics[width=0.75\columnwidth]{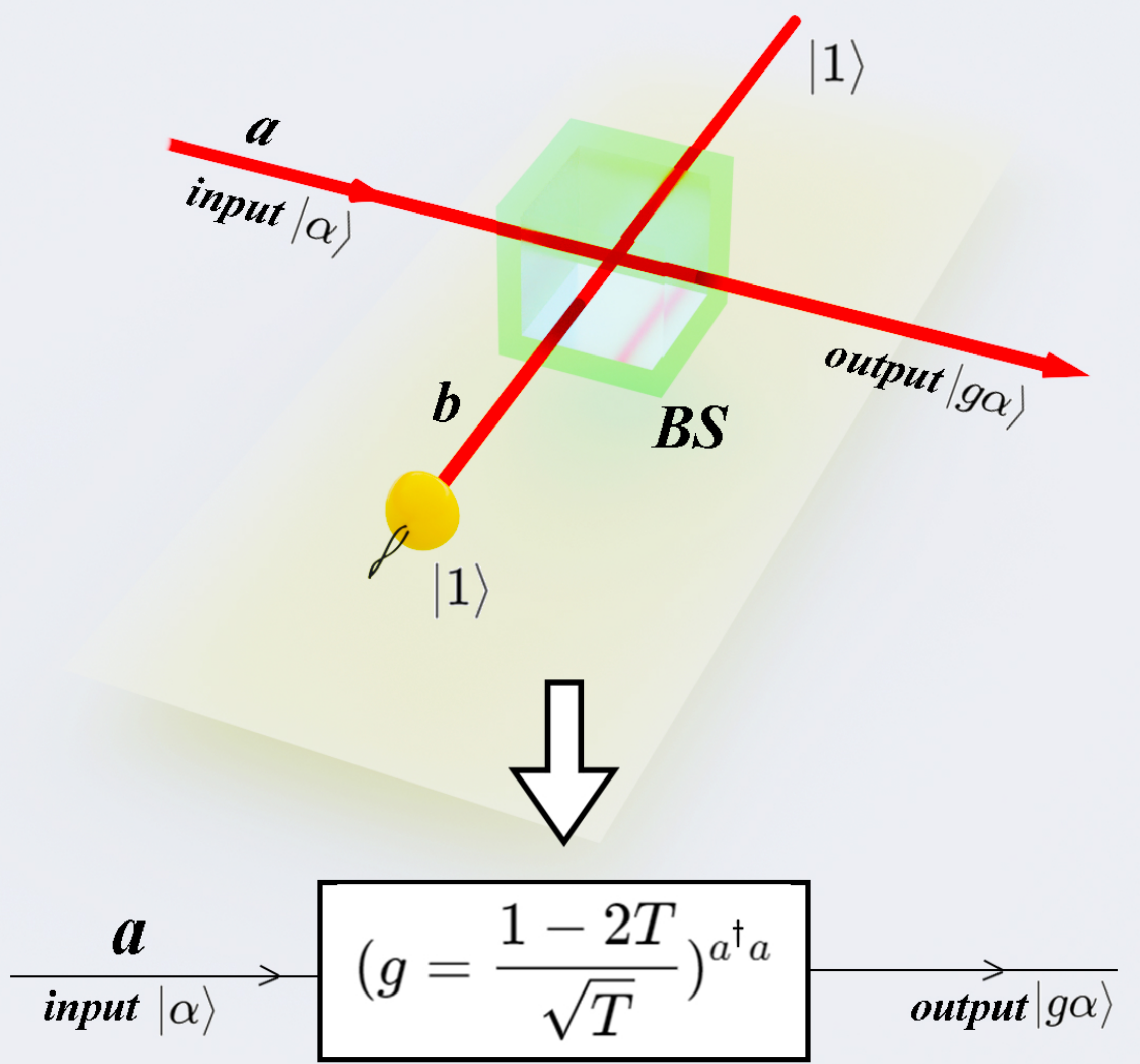}
\caption{{}(Color online) The realistic NLA process via the single-photon
catalysis where the input single photon $\left\vert 1\right\rangle $ in mode
$b$ interferes with a beam splitter (BS) with the modulated transmissivity $%
T $, and finally is registered by the single-photon counter in the
corresponding output port. When the single-photon catalysis acts on an input
small coherent state $\left\vert \protect\alpha \right\rangle $ in mode $a$,
the output state can be described as $\left\vert g\protect\alpha %
\right\rangle $ with the gain factors $g=$\ $(1-2T)/\protect\sqrt{T}$. }
\end{figure}

Through the above two sections, we have obtained two general frameworks of
the QZZB with the NLA-based CS in noisy scenarios, including the photon loss
and the phase diffusion. Under this circumstance, here we shall respectively
derive two novel and tighter Heisenberg error limits with the NLA-based CS
in terms of the photon loss and the phase diffusion.

Let $\left\vert \psi _{S}\right\rangle $=$\sum_{n=0}^{\infty
}c_{n}\left\vert n\right\rangle $ be a purification of the initial quantum
state, where each $\left\vert n\right\rangle $ is an eigenvector of the
photon number operator $\hat{n}$ with an eigenvalue $n$. Thus, based on Eq. (%
\ref{10}), the corresponding lower bound of the fidelity for photon-loss
scenarios can be represented as%
\begin{equation}
F_{Q_{1}}\left( \tau \right) =\left\vert \sum_{n=0}^{\infty }\left\vert
c_{n}\right\vert ^{2}\left[ \eta e^{-i\tau }+\left( 1-\eta \right) e^{i\tau
\mu _{1}}\right] ^{n}\right\vert ^{2}.  \label{26}
\end{equation}%
For sake of calculations, now let us consider two extreme cases, i.e., small
($\eta \longrightarrow 1$) and large ($\eta \longrightarrow 0$) photon
losses. In these conditions, we can respectively obtain the following two
inequalities%
\begin{equation}
F_{Q_{1}}\left( \tau \right) \geq \ddot{F}_{Q_{1}}\left( \tau \right)
=\left\{
\begin{array}{c}
Z_{1},(\eta \longrightarrow 1) \\
Z_{2},(\eta \longrightarrow 0)%
\end{array}%
\right. ,  \label{27}
\end{equation}%
where $Z_{1}=\left\vert \text{Tr}[\hat{\rho}_{S}\eta ^{\hat{n}}e^{-i\hat{n}%
\tau }[1+\tilde{\eta}e^{i\tau \left( \mu +1\right) }\hat{n}]]\right\vert
^{2},$ $Z_{2}=\left\vert \text{Tr}[\hat{\rho}_{S}\left( 1-\eta \right) ^{%
\hat{n}}e^{i\tau \mu _{1}\hat{n}}\left[ 1+\left. e^{-i\tau \left( \mu
_{1}+1\right) }\hat{n}\right/ \tilde{\eta}\right] ]\right\vert ^{2}$ and $%
\tilde{\eta}$=$(1-\eta )/\eta .$ It is worthy mentioning that $\ddot{F}%
_{Q_{1}}\left( \tau \right) $ is a lower bound of the $F_{Q_{1}}\left( \tau
\right) ,$ which can be only used for $\eta \longrightarrow 1$ and $\eta
\longrightarrow 0$.

Thus, by substituting Eq. (\ref{27}) into Eq. (\ref{12}), we can obtain the
Heisenberg error limit of the QZZB in the photon losses, which is given by%
\begin{equation}
\sum\nolimits_{L_{1}}\geq \sum\nolimits_{H}^{PL}=\frac{\pi }{8}%
\int_{0}^{2\pi }\max_{\mu }\left[ \ddot{F}_{Q_{1}}\left( \tau \right) \right]
\sin (\tau /2)d\tau .  \label{28}
\end{equation}%
To facilitate the elaborations of the compactness in Sec. V, the $%
\sum\nolimits_{H}^{PL}$ can be described as the PL-type bound. In addition,
Y. R. Zhang and H. Fan invoked the quantum speed limit theorem to derive two
kinds of Heisenberg error limits, i.e., the ML-type bound and the MT-type
bound, in both the photon loss and the phase diffusion \cite{36}. For
photon-loss scenarios, these two kinds of Heisenberg error limits can be
given by
\begin{equation}
\sum\nolimits_{L_{1}}\geq \max \left[ \frac{c_{ML}}{\eta ^{2}\left\langle
\hat{n}\right\rangle ^{2}},\frac{c_{MT}}{\frac{\eta \left\langle \hat{n}%
\right\rangle \Delta \hat{n}^{2}}{(1-\eta )\Delta \hat{n}^{2}+\eta
\left\langle \hat{n}\right\rangle }}\right] ,  \label{29}
\end{equation}%
where $\left\langle \hat{n}\right\rangle $=$\left\langle \psi
_{S}\right\vert \hat{n}\left\vert \psi _{S}\right\rangle $ and $\Delta \hat{n%
}^{2}$=$\left\langle \psi _{S}\right\vert \hat{n}^{2}\left\vert \psi
_{S}\right\rangle -\left\langle \psi _{S}\right\vert \hat{n}\left\vert \psi
_{S}\right\rangle ^{2}$ are the average value and variance of the photon
number operator $\hat{n},$ $c_{ML}$=$1/(80\lambda ^{2})$ and $c_{MT}$=$\pi
^{2}/16-1/2$ with $\lambda \approx 0.7246.$ Based on the Eqs. (\ref{28}) and
(\ref{29}), we can get the more tightness Heisenberg error limit for the
photon loss scenario, i.e.,%
\begin{equation}
\sum\nolimits_{L_{1}}\geq \max \left[ \frac{c_{ML}}{\eta ^{2}\left\langle
\hat{n}\right\rangle ^{2}},\frac{c_{MT}}{\frac{\eta \left\langle \hat{n}%
\right\rangle \Delta \hat{n}^{2}}{(1-\eta )\Delta \hat{n}^{2}+\eta
\left\langle \hat{n}\right\rangle }},\sum\nolimits_{H}^{PL}\right] .
\label{30}
\end{equation}%
When inputting the NLA-based CS $\hat{\rho}_{S}=\left\vert \gamma
\right\rangle \left\langle \gamma \right\vert $, we can respectively\ derive
Eq. (\ref{27}) as%
\begin{equation}
\ddot{F}_{Q_{1}}\left( \tau \right) _{\gamma }=\left\{
\begin{array}{c}
Ae^{B_{0}}\left( 1+C_{0}+D_{0}\right) ,(\eta \longrightarrow 1) \\
Ae^{C_{0}}\left( 1+B_{0}+D_{1}\right) ,(\eta \longrightarrow 0)%
\end{array}%
\right. ,  \label{31}
\end{equation}%
with%
\begin{eqnarray}
A &=&\exp \left( -2|\gamma |^{2}\right) ,  \notag \\
B_{0} &=&2\eta |\gamma |^{2}\cos \tau ,  \notag \\
C_{0} &=&2\left( 1-\eta \right) |\gamma |^{2}\cos \left( \tau \mu \right) ,
\notag \\
D_{0} &=&\left( 1-\eta \right) ^{2}|\gamma |^{4},  \notag \\
D_{1} &=&\eta ^{2}|\gamma |^{4}.  \label{32}
\end{eqnarray}%
Finally, by substituting Eq. (\ref{31}) into Eq. (\ref{28}), we can obtain
the PL-type bound with the NLA-based CS in the photon losses.

On the other hand, based on Eq. (\ref{21}), we can also derive another lower
bound of the $F_{Q_{2}}\left( \tau \right) $ for the phase diffusion, which
can be expressed as%
\begin{equation}
F_{Q_{2}}\left( \tau \right) \geq \ddot{F}_{Q_{2}}\left( \tau \right)
=\varphi (\kappa ,\tau ,\mu )\left[ 1-2\lambda \tau \left\langle \hat{n}%
\right\rangle \left\vert \mu -1\right\vert \right] .  \label{33}
\end{equation}%
Based on Eq. (\ref{33}), the corresponding Heisenberg error limit in the
phase diffusion can be given by
\begin{equation}
\sum\nolimits_{L_{2}}\geq \sum\nolimits_{H}^{PD}=\frac{\pi }{8}%
\int_{0}^{2\pi }\max_{\mu }\left[ \ddot{F}_{Q_{2}}\left( \tau \right) \right]
\sin (\tau /2)d\tau .  \label{34}
\end{equation}%
Similarly, for convenience, the $\sum\nolimits_{H}^{PD}$ can be described as
the PD-type bound. As a matter of fact, Ref. \cite{17} also gives the other
two types of Heisenberg error limits in the phase diffusion scenario, i.e.,%
\begin{equation}
\sum\nolimits_{L_{2}}\geq \max \left[ \frac{c_{ML}}{\min\limits_{\mu
}\left\langle \hat{H}\right\rangle _{+}^{2}},\frac{c_{MT}}{\min\limits_{\mu
}\Delta \hat{H}^{2}}\right] ,  \label{35}
\end{equation}%
where $\left\langle \hat{H}\right\rangle _{+}$=$\left\langle \hat{n}%
\right\rangle \left\vert 1-\mu \right\vert +\left. \left\vert \mu
\right\vert \right/ 2\sqrt{2\pi }\kappa $ and $\Delta \hat{H}^{2}$=$\Delta
\hat{n}^{2}\left( 1-\mu \right) ^{2}+\mu ^{2}/8\kappa ^{2}.$ According to
Eqs. (\ref{34}) and (\ref{35}), we can define the more tightness Heisenberg
error limit for phase-diffusion scenarios, i.e.,%
\begin{equation}
\sum\nolimits_{L_{2}}\geq \max \left[ \frac{c_{ML}}{\min\limits_{\mu
}\left\langle \hat{H}\right\rangle _{+}^{2}},\frac{c_{MT}}{\min\limits_{\mu
}\Delta \hat{H}^{2}},\sum\nolimits_{H}^{PD}\right] .  \label{36}
\end{equation}%
For the NLA-based CS $\hat{\rho}_{S}=\left\vert \gamma \right\rangle
\left\langle \gamma \right\vert $, likewise, we can rewrite the Eq. (\ref{33}%
) as%
\begin{equation}
\ddot{F}_{Q_{2}}\left( \tau \right) _{\gamma }=\varphi (\kappa ,\tau ,\mu
)[1-2\lambda \tau \left\vert \mu -1\right\vert |\gamma |^{2}].  \label{37}
\end{equation}%
Thus, by substituting Eq. (\ref{37}) into Eq. (\ref{34}), we can achieve the
PD-type bound with the NLA-based CS in phase-diffusion scenarios.

\section{The performance analyses of the QZZB with the NLA-based CS in noisy
scenarios \ }

\begin{figure}[tbp]
\label{Fig7} \centering \includegraphics[width=\columnwidth]{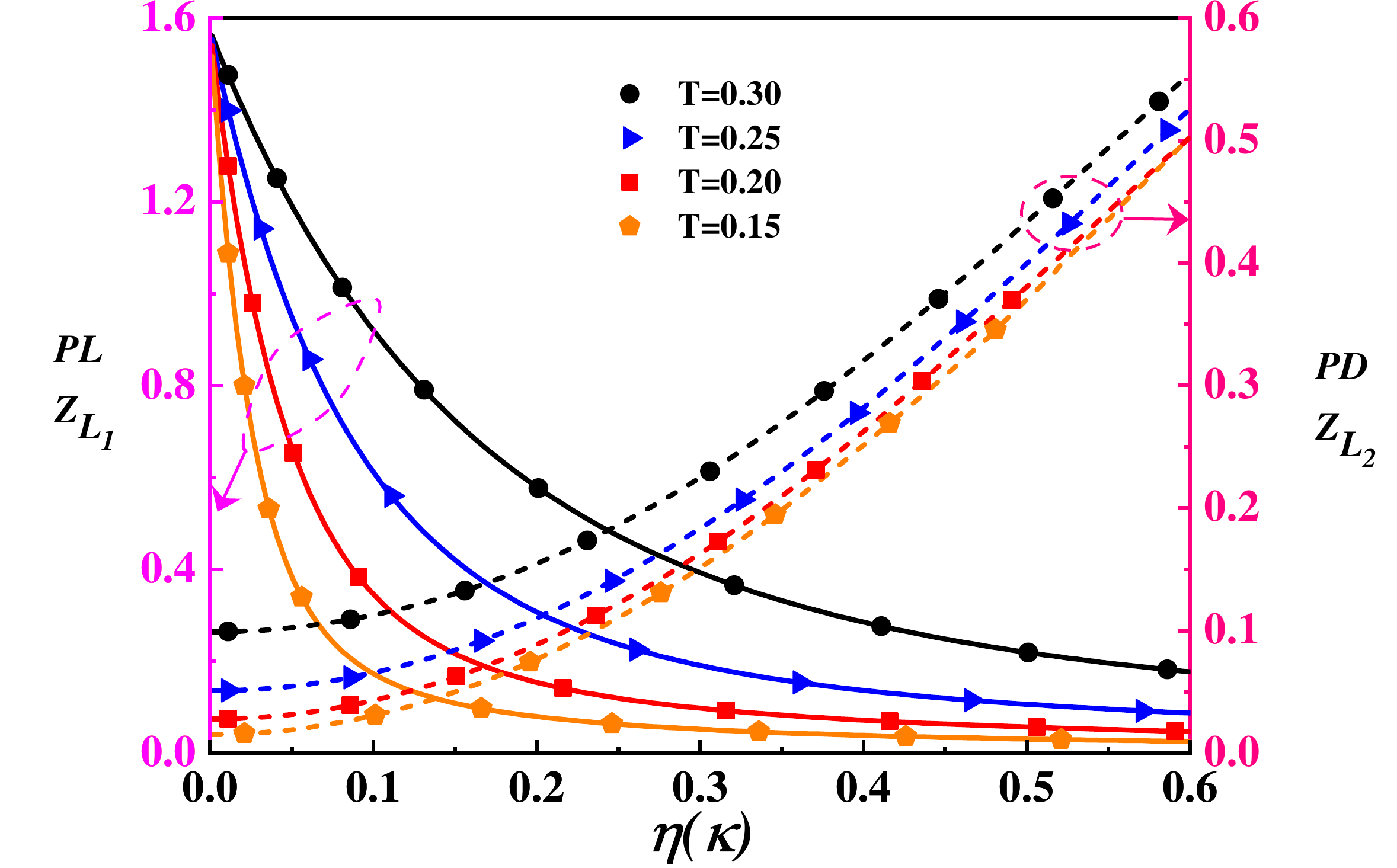}
\caption{{}(Color online) The effects of the single-photon catalysis on the
QZZB in the photon-loss (left ordinate) and the phase-diffusion (right
ordinate) cases changing with $\protect\eta $ and $\protect\kappa $ for
several different modulated transmissivities $T$ $=$ $0.3,0.25,0.2$ and $%
0.15 $, when given $N=4$. }
\end{figure}
\begin{figure*}[tbp]
\label{Fig8} \centering%
\subfigure[]{
\centering
\includegraphics[width=0.95\columnwidth]{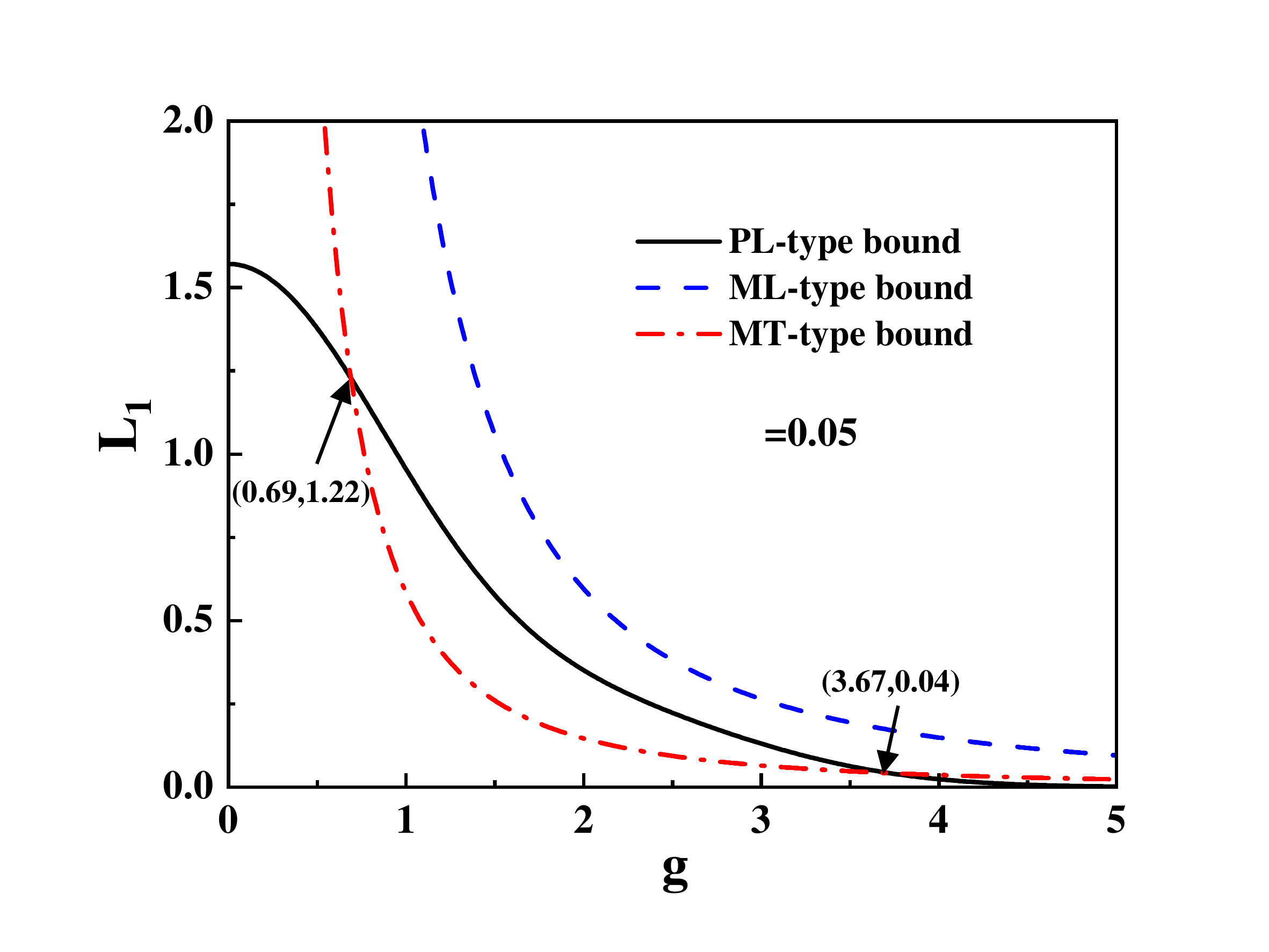}
\label{Fig8a}
}
\subfigure[]{
\includegraphics[width=0.96\columnwidth]{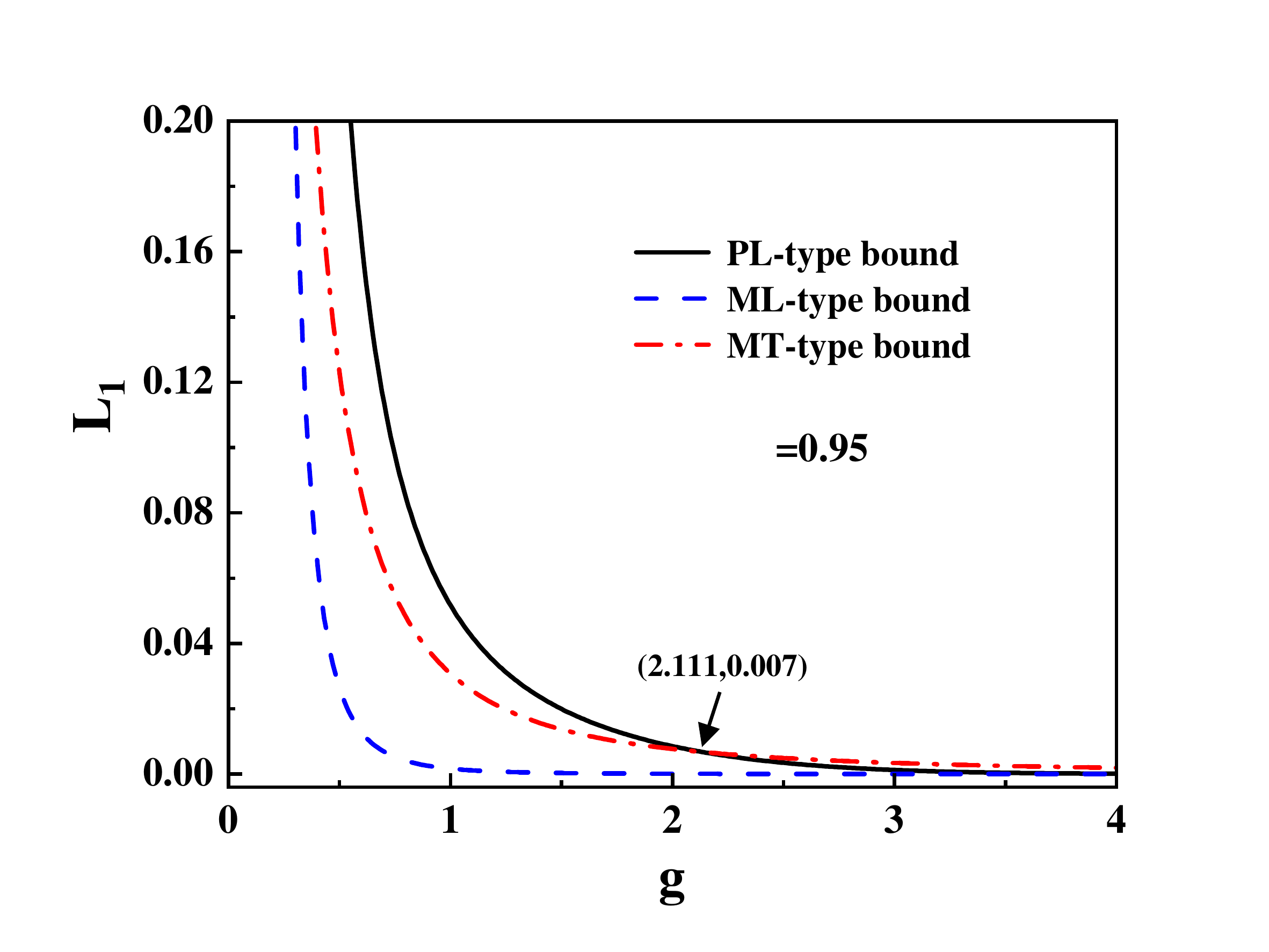}
\label{Fig8b}
}
\caption{{}(Color online) For photon-loss cases, i.e., (a) $\protect\eta %
=0.05$ and (b) $\protect\eta =0.95$, the QZZB as a function of the gain
factor $g$ at a fixed $N=4$ for several different bounds, involving the
PL-type bound, the ML-type bound and the MT-type bound.}
\end{figure*}
\begin{figure}[tbp]
\label{Fig9} \centering\includegraphics[width=0.95\columnwidth]{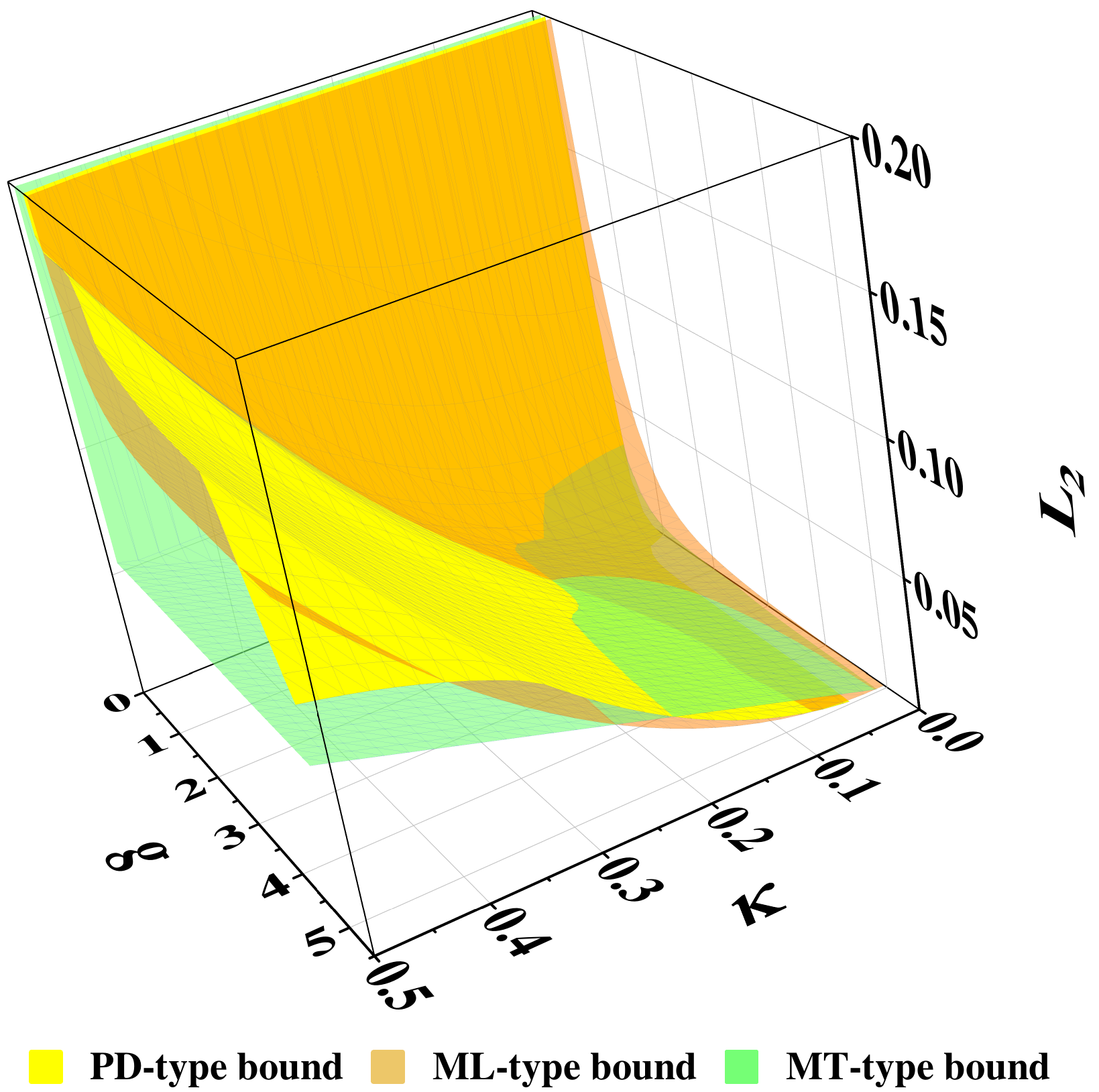}
\caption{{}(Color online) For photon-diffusion cases, the QZZB as a function
of the gain factor $g$ \ and the photon-diffusion strength $\protect\kappa $
at a fixed $N=4$ for several different bounds, involving the PD-type bound
(yellow surface), the ML-type bound (orange surface) and the MT-type bound
(green surface). }
\end{figure}
Now, let us present the phase estimation performance of the QZZB in noisy
scenarios, together with the NLA-based CS, including the ideal and feasible
NLA cases. For this purpose, we first plot the QZZB for different ideal NLA
gain factor of $g=1,2,3$ changing with the average photon number $N$ of the
initial CS when given photon-loss strengths $0.5$ (dashed lines) and $\eta
=1 $ (solid lines), as pictured in Fig. 2. Note that for comparison, $g=1$
corresponding to the QZZB without the NLA effects is also presented (black
lines). We can see that with the increase of $g=1,2,3,$ the phase estimation
performance of the QZZB can be improved effectively in the absence ($\eta =1$%
) and presence $(\eta =0.5)$ of photon losses, which means that the
introduction of ideal NLA can significantly reduce the phase estimation
uncertainty even with photon losses. To explain this phenomenon, when given
an average photon number $N=4$, we show the photon-loss generalized fidelity
$\tilde{F}_{1}(\tau )$ changing with $\tau $ at the same parameters, as
pictured in Fig. 3.\ Obviously, based on Eq. (\ref{12}), the area enclosed
by horizontal coordinate and curve line indicates the value of the QZZB,
implying that the smaller the area, the better the phase estimation
performance of the QZZB. It is clearly seen from Fig. 3 that the area with
and without photon-loss scenarios can be further reduced by increasing the
gain factors $g=1,2,3$. In addition, at a fixed ideal NLA, we also consider
the effects of more general photon-loss strength and ideal NLA on the QZZB,
as shown in Fig. 4(a). At the same gain factor of ideal NLA, the phase
estimation performance of the QZZB is getting worse and worse as $\eta $
decreases. Even so, we can improve this phase estimation performance by
increasing the gain factor $g$. In particular, the more serious the photon
losses, the more obvious the improvement effect, as pictured in Fig. 4(b)
where we plot the differences of the QZZB between without and with the ideal
NLA, i.e., $\sum\nolimits_{Z_{L_{1}}\alpha }^{PL}-$ $\sum%
\nolimits_{Z_{L_{1}}\gamma }^{PL}$.\

On the other hand, we further analyze the phase estimation performance of
the QZZB with the ideal NLA-based CS in phase-diffusion scenarios. Thus, we
show the corresponding QZZB as a function of the average photon number $N$
and the photon-diffusion strength $\kappa $ for several different gain
factors $g=1,2,3$, as pictured in Fig. 5(a). We can easily see that the
value of the QZZB can be decreased by the increase of $N$ or the decrease of
$\kappa $. Particularly, compared to without the gain factor $g=1$ (black
lines), the corresponding QZZB can be further decreased by increasing the
gain factors $g=2,3$, which means that the NLA can effectively improve the
phase estimation performance of QZZB even in the presence of phase
diffusion. The reason for this phenomenon is that with the increase of $%
g=1,2,3$, the area of the phase-diffusion generalized fidelity $\tilde{F}%
_{2}(\tau )$ enclosed by horizontal coordinate and curve line can be further
reduced, as shown in Fig. 5(b).

From the viewpoint of realizing NLA, the single-photon catalysis is an
alternative scheme, as showed in Fig. 6, where $\left\vert \alpha
\right\rangle \rightarrow \left\vert g\alpha \right\rangle $ with the gain
factor $g=$\ $(1-2T)/\sqrt{T}$ if the single-photon catalysis with the
modulated transmissivity $T$ works on a small input coherent state $%
\left\vert \alpha \right\rangle $. In order to see the effects of the
single-photon catalysis on the QZZB in noisy scenarios, at a fixed $N=4$, we
also consider the QZZB changing with $\eta $ and $\kappa $ for several
different values of $T$ $=$ $0.3,0.25,0.2$ and $0.15$, as pictured in Fig.
7. We can see that with the decrease of $T=0.3,0.25,0.2,0.15$, the phase
estimation performance of the QZZB in photon-loss and phase-diffusion
scenarios can be enhanced dramatically, which results from that when $T<1/4$%
, the corresponding gain factor $g$ greater than $1$ has the effect of the
NLA.

Finally, we examine the compactness of several bounds with the NLA-based CS
in noisy scenarios, including the PL-(or PD-)type bound, the ML-type bound
and the MT-type bound \cite{36}. For the photon-loss cases, i.e., $\eta
=0.95 $ and $0.05,$ when given $N=4$, we make comparisons about the
compactness of the aforementioned bounds changing with the NLA gain factor $%
g $, as pictured in Fig. 8. When given a strong photon loss $\eta =0.05$,
for all the bounds, the ML-type bound has the best compactness.
Additionally, the compactness of our PL-type bound only perform better than
that of the MT-type bound when $g\in $\{$0.69,3.67$\}. More importantly, for
a small photon loss $\eta =0.95$, we can adjust the gain factor $g$ to less
than $2.11$ to obtain that our PL-type bound shows the best compactness.
Different from the above photon-loss cases, by adjusting the value range of
the gain factor $g$, the compactness of our PD-type bound (yellow surface)
can show better compactness only in the case of strong phase diffusion
(e.g., $\kappa =0.25$). These results fully indicate that the tighter phase
precision limit can be obtained by taking appropriate values of the gain
factor $g$ especially in small photon-loss and strong phase-diffusion
scenarios.

\section{Conclusions}

In this paper, we have derived the general form of the QZZB for phase
estimation with the NLA-based CS in noisy scenarios, involving the photon
loss and the phase diffusion, by using the variational method. Under these
circumstances, more importantly, we have also obtained Heisenberg error
limits of the QZZB for phase estimation, which can be called as the PL-type
bound in the photon loss and the PD-type bound in the phase diffusion. Our
results show that for both photon-loss and phase-diffusion scenarios, the
corresponding phase estimation performance of the QZZB can be enhanced
effectively by increasing the NLA gain factor. Particularly, the more
serious the photon losses, the more obvious the improvement effect. In
addition, we have also considered the effects of realistic operational NLA,
i.e., the single-photon catalysis, on the phase estimation performance of
the QZZB. It is found that such estimation performance in photon-loss and
phase-diffusion scenarios can be enhanced dramatically by regulating the
transmissivity of BS less than $1/4$.\ Finally, we have examined the
compactness of several bounds with the NLA-based CS in noisy scenarios,
including the PL-(or PD-)type bound, the ML-type bound and the MT-type
bound. Our results show that the tighter phase precision limit can be
obtained by taking appropriate values of the NLA gain factor especially for
small photon-loss and strong phase-diffusion cases. Through these studies,
our findings demonstrate the significant benefits of NLA techniques in
quantum metrology in noisy environments, and show its promising applications
in a wide range of fields in quantum science and technology.

\section*{Acknowledgments}

This work was supported by the China Scholarship Council; National Nature
Science Foundation of China (Grant No. 62161029); Natural Science Foundation
of Jiangxi Province (Grant No. 20202BABL202002). Wei Ye is supported by both
Jiangxi Provincial Natural Science Foundation (CA202304328) and Scientific
Research Startup Foundation (Grant No. EA202204230) at Nan chang Hangkong
University.

\end{document}